\documentstyle[amssymb,amsfonts,psfig]{elsart}
\begin{document} \begin{frontmatter}
\title{Optimal quantum estimation of the coupling between two 
bosonic modes} 
\author[mauro]{G. Mauro D'Ariano} \author[matteo]{Matteo G. A. Paris}
\author[paolo]{Paolo Perinotti} \address[mauro]{Sezione INFN, 
Universit\'a di Pavia, via Bassi 6, I-27100 Pavia, ITALIA}
\address[matteo]{Unit\'a INFM, Universit\'a di Pavia, via Bassi 6, 
I-27100 Pavia, ITALIA} \address[paolo]{Sezione INFN, Universit\'a di 
Milano, via Celoria 16, I-20133 Milano, ITALIA}
\date{\today}\maketitle
\begin{abstract}
We address the problem of the optimal quantum estimation of the 
coupling parameter of a bilinear interaction, such as the 
transmittivity of a beam splitter or the internal phase-shift of an 
interferometer. The optimal measurement scheme exhibits Heisenberg 
scaling of the measurement precision versus the total energy.
\end{abstract}
\end{frontmatter}
\section{Introduction}
How effectively may we estimate the strength of a simple interaction 
such as the transmittivity of a beam splitter or the phase-shift imposed
in the internal arms of an interferometer ? The Hamiltonian describing the 
bilinear coupling between two bosonic modes has the form
\begin{equation}
H=\kappa(a^{\dag}b+b^{\dag}a)
\label{eq:hamilt}\:,\end{equation}
where $\kappa$ depends on the specific interaction under consideration. 
By using the Schwinger representation of the $SU(2)$ Lie algebra, with 
generators
\begin{equation} 
J_x =\frac12(a^{\dag}b+b^{\dag}a)\:, \quad J_y =
\frac{1}{2i}(a^{\dag}b-b^{\dag}a)\:, \quad J_z =
\frac12(a^{\dag}a-b^{\dag}b) \:, \end{equation}
we can rewrite the Hamiltonian as $H= 2\kappa  J_x$ and the evolution
operator as 
\begin{equation}
U_\psi =\exp (-{\rm i} J_x \psi) 
\label{evol}\:,
\end{equation}
where the global coupling constant $\psi$ is equal to $2\kappa\Delta t$,
$\Delta t$ being the effective interaction time. The evolution in Eq.
(\ref{evol}) describes, for example, the interaction of two light
modes in a beam splitter with transmittivity $\tau=\cos^2 \psi$ 
\cite{man}, or, apart from a fixed rotation, the evolution of the 
arm modes in a Mach-Zehnder interferometer, with $\psi$
representing the phase-shift between arms \cite{mil,vis}. 
If the initial preparation of two modes is described by the density 
matrix $\rho_0$ the evolved state in the interaction picture is given 
by
\begin{equation}
\rho_{\psi}=\exp (-{\rm i} J_x \psi)\rho_0\exp ({\rm i} J_x \psi)\:.
\end{equation}
In this paper we devote our attention to the estimation of $\psi$ through
measurements performed on $\rho_{\psi}$. We denote the generic POVM for such
estimation process by ${\rm d}\nu(\phi)$, such that the outcomes of the 
measurement are distributed according to
\begin{equation}
p(\phi\vert\psi){\rm d}\phi  ={\rm Tr}[\rho_{\psi}
{\rm d}\nu (\phi)]\:,
\label{eq:dispro}
\end{equation}
where $p(\phi\vert\psi)$ represents the conditional probability of 
registering the outcome $\phi$ when the true value of the parameter is 
$\psi$.  \par
Our objective is to find the best strategy, {\em i.e.} the POVM that provides
the optimal estimation of the parameter $\psi$ \cite{hel}. Since $\psi$ is 
manifestly a phase-shift, we can use general results from phase estimation 
theory, which provides the optimal POVM to estimate the phase-shift 
induced by a {\em phase generator}, {\em i.e.} a selfadjoint operator 
with discrete, equally spaced, spectrum. 
The optimality criterion is given in terms of the minimization of the 
mean value of a cost function \cite{hel} that assesses the quality of 
the strategy, {\em i.e.} it weights the errors in the estimates. Since a 
phase-shift is a $2\pi$-periodic parameter, the cost function must be
a $2\pi$-periodic even function of $(\psi-\phi)$, {\em i.e.} 
it has Fourier expansion with cosines only. The appropriate concavity
properties of the cost function are reflected by expansion coefficients 
which are all negative apart the irrelevant additive constant.
A cost function with such Fourier expansion has been firstly considered by
Holevo \cite{hol}, and for such reason it is usually referred to as 
belonging to Holevo's class. Notice that the optimal POVM for a given 
state $\rho_0$ is the same for every cost function in this class. 
A more general quantum estimation approach for different kinds of 
phase-shift and general quantum system is given in Ref. \cite{gpe}.
There, it is also shown that the kind of problem we are presently 
dealing with is, in a certain sense, the best situation, since the 
spectrum of our phase generator is the whole set of integers 
${\Zset}$, 
including negative ones. In this case, there is an
optimal orthogonal projective POVM, which can be regarded as the spectral 
resolution of a self-adjoint phase operator. 
However, if the estimation is performed with the constraint of bounded or
fixed energy, the optimal POVM and the optimal input state $\rho_0$ do not
correspond to a canonical quantum observable scheme \cite{cano}.
Moreover, in general the optimal POVM depends on the input state, and the 
optimal POVM of Ref. \cite{hol,hel} holds only for pure states, whereas 
a generalization to a class of mixed states, the so-called {\em phase-pure}
states, has been considered in Ref. \cite{gpe,iso}.  \par
The optimal POVM, in the sense described above, provides an unbiased
estimation of $\psi$ for preparation $\rho_0$ of the two modes, in formula
$$\langle \phi\rangle =\psi \quad {\rm with}\quad \langle \phi\rangle=
\int_0^{2\pi}\phi\:{\rm Tr}[\rho_\psi\:{\rm d}\nu(\phi)]
\qquad\forall \rho_\psi = U_\psi \rho_0 U^\dag_\psi
\:.$$ On the other hand, we want also to find the optimal state $\rho_0$ 
for the estimation of $\psi$ according to some cost function, which quantifies 
the noise of the estimation.
The customary root mean square is not a good choice for a cost function, 
since the function $(\phi-\psi)^2$ is not $2\pi$-periodic.
A good definition for the precision of the measurement is given by the 
average of the cost function $C(\phi-\psi)=4\sin^2 (\frac{\phi-\psi}{2})$, 
{\em i. e.} a "periodicized variance", which obviously belongs to the 
Holevo's class. If the estimates occur within a small interval around
the true value of the parameter $\psi$, one has approximately 
$C(\phi-\psi)\simeq(\phi-\psi)^2$, whence 
$\delta\psi=\sqrt{C}$ can be assumed as a reasonable measure of the 
precision of the measurement.
\section{Optimal estimation of the coupling parameter}
In order to solve our estimation problem, let us consider the 
following unitary transformation
\begin{equation}
{\cal U} =\exp \left\{-\frac{\pi}{4}(a^{\dag}b-b^{\dag}a)\right\}=
\exp \left\{-{\rm i}\frac{\pi}{2} J_y\right\}\label{eq:unitransf}\:.
\end{equation}
Using Eq. (\ref{eq:unitransf}) we may rewrite Eq. (\ref{eq:dispro}) 
in the more familiar form of rotation along the z-axis
\begin{eqnarray}
p(\phi\vert\psi){\rm d}\phi &=& {\rm Tr}\left[{\cal U}\rho_{\psi}{
\cal U}^{\dag}\:{\cal U}{\rm d}\nu (\phi){\cal U}^{\dag}\right] 
\nonumber \\ &=& {\rm Tr}\left[
\exp(-{\rm i} J_z \psi){\cal U}\rho_0{\cal U}^{\dag} \exp({\rm i} J_z \psi)\:
{\cal U}{\rm d} \nu(\phi){\cal U}^{\dag}\right] \nonumber \\ &=&
{\rm Tr}\left[\exp(-{\rm i} J_z \psi)
R_0\exp({\rm i} J_z \psi)\: {\rm d}\mu(\phi)\right] 
\:,\label{eq:disp1}
\end{eqnarray}
where we used the identity ${\cal U}J_x{\cal U}^{\dag}=J_z$. Equation
(\ref{eq:disp1}) shows that the problem of estimating the shift generated by
$J_x$ on the state $\rho_0$ is equivalent to that of estimating the same shift
generated by $J_z$ on the rotated state $R_0={\cal U}\rho_0{ \cal U}^{\dag}$.
In particular, any POVM ${\rm d}\nu(\phi)$ to estimate the $J_x$-induced shift
can be written as $d\nu(\phi)={\cal U}^{\dag}{\rm d}\mu(\phi){\cal U}$, where ${\rm
d}\mu(\phi)$ is a POVM for the $J_z$-induced shift estimation.  \par
For pure states $R_0=\vert\psi_0\rangle\rangle\langle\langle\psi_0\vert$ 
(in the following we use double brackets for two-mode vectors) the
degeneration of the spectrum of $a^{\dag}a-b^{\dag}b$ can be treated 
using the technique introduced in \cite{gpe}, and the optimal
POVM for cost functions in Holevo's class \cite{hol} 
is proved to be of the form
\begin{equation}
{\rm d}\mu (\phi)=\frac{{\rm d}\phi}{2\pi}\:
\vert E_\phi\rangle\rangle\langle\langle E_\phi\vert
\end{equation}
with the vectors $|E_\phi\rangle\rangle$ given by
\begin{equation}
\vert E_\phi\rangle\rangle =\sum_{d\in{\mathbb Z}}{\rm e}^{i d\phi}\:
\vert d\rangle\rangle\:. 
\end{equation}
The vectors $\vert d\rangle\rangle$ are certain eigenvectors of 
$D=a^{\dag}a-b^{\dag}b$ built by picking up, in every eigenspace 
${\cal  H}_d$ of the eigenvalue $d$, the normalized vector parallel to
the projection of $\vert\psi_0\rangle \rangle$ on ${\cal H}_d$. 
In order to be more specific, let us consider an input state of the form
\begin{equation}
\vert\psi_0\rangle
\rangle=\sum_{n=0}^{\infty}\psi_n ^{(0)}\vert n,0\rangle
\rangle+\sum_{n=0}^{\infty}\sum_{d=1}^{\infty}
\left[\psi_n ^{(d)}\vert n,d\rangle
\rangle+\psi_n ^{(-d)}\vert n,-d\rangle
\rangle\right]\:, \end{equation}
where $\vert n,d\rangle \rangle$ is given by 
\begin{eqnarray}
\vert n,d\rangle \rangle \equiv \left\{
\begin{array}{cc}
\vert n+d \rangle_a\vert n  \rangle_b  & {\rm if}\:\: d\geq 0  \\
\vert n   \rangle_a\vert n-d\rangle_b  & {\rm if}\:\: d<0
\end{array}\right.
\label{defnd}\:.
\end{eqnarray}
The projection of $\vert\psi_0\rangle \rangle$ in ${\cal H}_d$ is equal to
\begin{equation}
\sum_{n=0}^{\infty}\psi_n^{(d)}\:\vert n,d\rangle\rangle\:,
\end{equation}
such that the eigenvector $\vert d\rangle \rangle$ reads as follows  
\begin{equation}
\vert d\rangle
\rangle =\frac{\sum_{n=0}^{\infty}\psi_n ^{(d)}\vert n,d\rangle
\rangle}{\sqrt{\sum_{n=0}^{\infty}\left|\psi_n ^{(d)}\right|^2}}\:,
\end{equation}
and the input state can be rewritten as 
$$\vert\psi_0 \rangle\rangle=\sum_{d\in{\mathbb Z}}\gamma_d\: \vert
d\rangle\rangle \qquad \gamma_d =\sqrt{\sum_{n=0}^{\infty}
\left|\psi_n^{(d)}\right|^2}\:.$$ 
Notice that the dependence of the POVM on the state
$\vert\psi_0\rangle\rangle$ is contained in the vectors 
$|E_\phi\rangle\rangle$.  \par
By adopting $C(\phi-\psi)$ as a cost function the average cost of the strategy
corresponds to the expectation value of the {\em cost} operator $C=2-{\rm E}_+
-{\rm E}_-$, where the raising and lowering operators $E_+$ and $E_-$ are
given by $${\rm E}_+= \sum_{d\in{\mathbb Z}}\vert d+1\rangle
\rangle\langle\langle d\vert \qquad {\rm E}_- ={\rm E}_+^{\dag}$$ (with the
vectors $\vert d\rangle \rangle$ defined as above).  \par
The optimization problem is that of minimizing the average cost of 
the strategy
\begin{eqnarray}
\bar{C} = \int_0^{2\pi}\frac{d\psi}{2\pi}\:\int_0^{2\pi}
\frac{d\phi}{2\pi}\: C(\phi-\psi)\: p(\phi|\psi) = 
{\rm Tr}\left[R_0\: C\right] \equiv
\langle\langle\psi_0\vert 2-{\rm E}_+ -{\rm E}_-\vert\psi_0\rangle\rangle
\nonumber \\\label{defC}\;,
\end{eqnarray}
with the constraint that the solution is a normalized state. The Lagrange 
function is given by
\begin{equation}
{\cal L}=\bar{C}-\lambda\langle\langle\psi_0\vert\psi_0\rangle
\rangle\;,
\label{eq:lagrange}
\end{equation}
with $\lambda$ being the Lagrange multiplier for the
normalization constraint.
The solution of this problem is a state with infinite mean energy
$N=\langle\langle\psi_0\vert a^{\dag}a+b^{\dag}b\vert\psi_0\rangle\rangle$.
In order to find physically realizable states, one must impose a constraint 
on $N$ too, and the Lagrange function becomes
\begin{equation}
{\cal L}=\bar{C}-\mu
\langle\langle\psi_0\vert a^{\dag}a+b^{\dag}b\vert\psi_0\rangle\rangle
-\lambda\langle\langle\psi_0\vert\psi_0\rangle\rangle\:,
\label{eq:lagrange2}
\end{equation}
$\mu$ being the Lagrange multiplier for the mean energy. It is useful, in
order to calculate the solution of this equation, to write the generic state
$\vert\psi_0\rangle\rangle$ in the following way
\begin{equation}
\vert\psi_0\rangle \rangle=\sum_{d\in{\mathbb Z}}\psi_d \sum_{n=0}^{\infty}c_{n,d}
\vert n,d\rangle\rangle\:.
\label{eq:decompdfix}
\end{equation}
The coefficients $c_{n,d}$ determine the normalized projection of
$\vert\psi\rangle \rangle$ into the eigenspace ${\cal H}_d$, whereas the
$\psi_d$'s are the coefficients which combine those projections.  \par
Using Eq. (\ref{eq:decompdfix}), the Lagrange function (\ref{eq:lagrange2}) 
explicitly shows terms accounting for the normalization of the vectors
$\sum_{n=0}^{\infty}c_{n,d}\vert n,d \rangle \rangle$ in each ${\cal H}_d$, 
with $\nu^{(d)}$ denoting Lagrange multipliers for the normalization
of projections, and rewrites as 
\begin{eqnarray}
{\cal L} & = & \sum_{d\in{\mathbb Z}}\left\{2\vert\psi_d
\vert^2\sum_{n=0}^{\infty} \vert c_{n,d}\vert^2-\left(\bar\psi_d
\sum_{n=0}^{\infty}\vert c_{n,d}\vert^2 \right)
\left(\psi_{d-1}\sum_{m=0}^{\infty}\vert
c_{m,d-1}\vert^2\right)+\right.\nonumber\\ & & -\left(\bar\psi_d
\sum_{n=0}^{\infty}\vert c_{n,d}\vert^2 \right)\left(\psi_{d+1}
\sum_{m=0}^{\infty}\vert c_{m,d+1}\vert^2 \right)+\nonumber\\ & & \left.
-\mu\vert\psi_d \vert^2 \sum_{n=0}^{\infty}\vert c_{n,d}\vert^2(2n+ \vert
d\vert)-\lambda\vert\psi_d \vert^2\sum_{n=0}^{\infty}\vert c_{n,d}\vert^2 -
\nu^{(d)}\sum_{n=0}^{\infty}\vert c_{n,d}\vert^2\right\}
\label{eq:lagrange3}\:.
\end{eqnarray}
By taking derivatives of the Lagrange function with respect to 
$c^*_{n,d}$ and $\psi^*_d$ with the constraints
\begin{eqnarray}
&&\sum_{n=0}^{\infty}\vert c_{n,d}\vert ^2 =1 \qquad \sum_{d\in{\mathbb Z}}
\vert\psi_d \vert^2 =1 \nonumber\\
&&\sum_{d\in{\mathbb Z}}\sum_{n=0}^{\infty}\vert\psi_d \vert^2 \vert c_{n,d}
\vert ^2 (2n+\vert d\vert)=N  \label{ccond}\:,
\end{eqnarray}
and by rephasing the $\vert d\rangle \rangle$'s we arrive at 
the system 
\begin{eqnarray}\left\{
\begin{array}{l}
(2-\lambda)\psi_d -\psi_{d-1}-\psi_{d+1}-\mu\sum_{n=0}^{\infty}(2n+\vert
d\vert) \vert c_{n,d}\vert^2 \psi_d=0 \\ \left[(2-\lambda)\psi_d^2 -2(\psi_d
\psi_{d-1}+\psi_{d+1} \psi_d )-\mu \psi_d^2(2n+\vert d\vert)-\nu^{(d)}\right] 
c_{n,d}= 0 \end{array} \right.
\label{eq:sistema2}
\end{eqnarray}
The second equation in (\ref{eq:sistema2}) implies that for a fixed $d$ only
one coefficient $c_{n,d}$ can be different from zero, say for the value $\bar
n$, and in this case $\vert c_{\bar n,d}\vert =1$
\footnote{\footnotesize
Indeed, let us suppose $c_{n,d}\neq 0$ for two values $n_1=m$ and $n_2=p$.
Then we must have $(2-\lambda)\psi_d^2 -2(\psi_d \psi_{d-1}+\psi_{d+1} \psi_d
)-\mu\psi_d^2(2p+\vert d\vert)-\nu^{(d)}=0$ which implies $2\mu
(m-p)\psi_d=0$. Since the case $\mu=0$ is not interesting and $\psi_d=0$ would
imply that the choice of the $c_{n,d}$ is completely arbitrary and irrelevant,
the only possibility is $m=p$.}. 
\par
The first equation of (\ref{eq:sistema2}) can therefore be rewritten as  
\begin{equation}
(2-\lambda)\psi_d -\psi_{d-1}-\psi_{d+1}-\mu(2n(d)+\vert d\vert) \psi_d=0,
\label{eq:formafinale}
\end{equation}
which allows us to obtain from the second one
\begin{equation}
\nu^{(d)}=-\psi_d (\psi_{d+1}+\psi_{d-1})\label{x2}\:.
\end{equation}
The solutions of Eq. (\ref{x2}) give local minima for the average cost 
$\bar{C}$, and one should solve the equation (\ref{eq:formafinale}) for 
arbitrary choices of $n(d)$,
looking for the optimal one. In the case $n(d)=0$ we have
\begin{eqnarray}
\frac{2(\lambda^{\prime}+\vert d\vert)}{\frac{2}{\mu^{\prime}}}
\psi_d =\psi_{d-1}+\psi_{d+1}\:,\label{23}
\end{eqnarray}
where $\mu^{\prime}=-\mu$ and
$\lambda^{\prime}=\frac{2-\lambda}{\mu^{\prime}}$.  Eq. \ref{23}
is the recursion equation for Bessel functions, with solution given by
\begin{equation}
\psi_d ={\cal N}^{-1/2}(\mu^{\prime},\lambda^{\prime})J_{\lambda^{\prime}+
\vert d\vert}\left(\frac{2}{\mu^{\prime}}\right),
\end{equation}
with ${\cal N}(\mu^{\prime},\lambda^{\prime})=\sum_{d\in{\mathbb Z}}
J_{\lambda^{\prime}+\vert d\vert}^2\left(\frac{2}{\mu^{\prime}}\right)$ 
and with the boundary conditions $J^{\prime}_{\lambda^{\prime}}
(2/\mu^\prime)=0$,
i.e. $J_{\lambda^{\prime}+1}=J_{\lambda^{\prime}-1}$.  Finally, to obtain the
optimal state $\rho_0$ one has to rotate $R_0=\vert\psi_0\rangle
\rangle\langle\langle\psi_0 \vert$ by the unitary transformation
(\ref{eq:unitransf}). \par 
\begin{figure}[h]
\centerline{\psfig{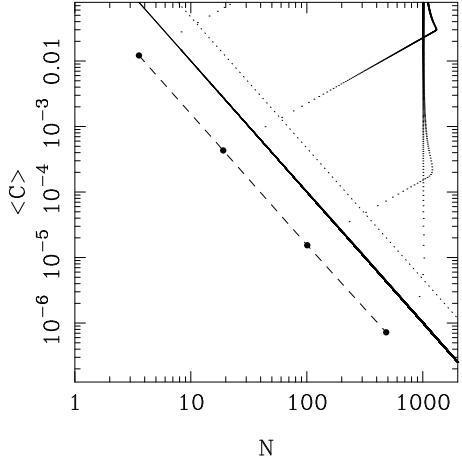}}
\caption{Average cost $\bar{C}$ as a function of the energy $N$ for the
optimal states (dashed line). Solid line is the function $1/N^2$. The points
above the solid line are other solutions of equation (\ref{eq:formafinale})
corresponding to local minima of the cost, distributed over lines
characterized by a fixed value of the Lagrange parameter $\mu$ (a single
dotted line is plotted, connecting points for different $\mu$ 
and increasing $N$).} \label{f:fig1} \end{figure}
In order to obtain the behaviour of the average cost versus the energy 
we numerically solved Eq. (\ref{eq:formafinale}) with $n(d)=0$. This problem
can be rewritten as the eigenvalue problem ${\bf A}\psi=\lambda\psi$ for the
matrix ${\bf A}$ with elements given by
\begin{equation}
({\bf A})_{m,n}=(2-\mu \vert m\vert)\delta_{m,n}-
\delta_{m,n+1}-\delta_{m,n-1}\:.
\end{equation}
Numerical diagonalization gives the power-law
$\bar{C}\simeq\frac{\gamma}{N^2}$ in the range $0\lesssim N
\lesssim 1000$ with $\gamma \simeq 0.1$.
This behaviour is plotted in Fig. \ref{f:fig1}, where also other solutions of
equation (\ref{eq:formafinale}) corresponding to local minima of the average
cost are shown. Since the phase distribution of the optimal state is singly
peaked we may also write $\delta\psi\simeq\sqrt{\bar{C}}\simeq
\gamma^{1/2}/N$, which means that the optimal states derived here 
are at the so-called Heisenberg limit of phase variance. 
\section{Conclusions}
In conclusion, we dealt with the problem of estimating the coupling 
constant of a bilinear interaction. In our approach 
the coupling constant, which appears in the exponent of the time 
evolution operator, has been treated as a phase parameter. 
The optimal POVM has been derived according to the 
theory of quantum phase estimation \cite{hol,hel,gpe,iso}. As noticed 
in Ref. \cite{mil} this resorts to an SU(2) estimation problem with 
the Schwinger two-mode boson realization. However, the representation 
is not irreducible, and a more complicated problem is faced. 
In this sense our results generalize those of Ref. \cite{mil}, where
the estimation of $SU(2)$ phase-shifts has been analyzed in irreducible 
subspaces: indeed we found an improved scaling of phase variance 
versus the total energy.
The degeneracy of the spectrum of the the Hamiltonian, can be treated
using the technique of Ref. \cite{gpe}, and, in this way, the problem is 
reduced to a nondegenerate one with spectrum ${\mathbb Z}$ for the 
phase-shift operator. \par
It is important to remark that the optimal POVM depends on the preparation
state. This is true for every kind of phase estimation problem, but in
the presence of degeneracy the dependence is crucial. In fact, one has to 
define the optimal POVM as a block diagonal operator, where the invariant 
subspaces are spanned by projections of the input state into the 
eigenspaces of the generator. From a practical point of view this 
means that optimal estimation of the phase-shift imposed to a state, needs
a measuring device which is adapted to the shifted state. 
We then optimized the input state by minimizing the average cost for fixed
input energy and found a power law $\delta\psi\simeq \gamma/N$ in a range 
$0\lesssim N \lesssim 1000$.\par
Notice that the law $N^{-1}$ is the same as in the optimal phase estimation
with only one mode \cite{rip}, however with a much smaller constant
$\gamma$ ($\gamma \simeq 0.1$ instead of $\gamma \simeq 1.36$). We think that
this phenomenon of improvement of phase sensitivity by increasing the number
of modes is the same considered in Ref. \cite{yue}, where an exponential
improvement versus the number of modes has been estimated when increasing the
number of modes and the the number of photons per mode, jointly in the same
proportion. 
\section*{Acknowledgement} 
This work has been cosponsored by MURST under the project 
"Quantum Information Transmission and Processing". The authors 
thank Dipartimento di Fisica "Alessandro Volta" for partial support. 


\begin{thebibliography}{99}
\bibitem{man} L. Mandel, E. Wolf {\em Optical Coherence and Quantum 
Optics} (Cambridge University Press, 1995).
\bibitem{mil} B. C. Sanders, G. J. Milburn, Phys. Rev. Lett. {\bf 75}, 2944
(1995).
\bibitem{vis} M. G. A. Paris, Phys. Rev. A {\bf 59}, 1615 (1999).
\bibitem{hol} A. S. Holevo, {\it Probabilistic and Statistical Aspects
of Quantum Theory} (North-Holland Publishing, Amsterdam, 1982).
\bibitem{hel} C.W.Helstrom, Found. Phys. {\bf 4}, 453 (1974); 
Int. J. Theor. Phys. {\bf 11}, 357 (1974); {\em Quantum Detection and 
Estimation Theory} (Academic Press, New York, 1976)
\bibitem{gpe}G. M. D'Ariano, C. Macchiavello, and M. F. Sacchi, 
Phys. Lett. A, {\bf 248} 103 (1998). 
\bibitem{cano} M. G. A. Paris, Nuovo Cimento B {\bf 111}, 1151 (1996).
\bibitem{iso} G. M. D'Ariano, C. Macchiavello, P. Perinotti, and 
M. F. Sacchi, Phys. Lett. A, {\bf 268} 241 (2000).
\bibitem{rip} G. M. D'Ariano e M. G. A. Paris, 
Phys. Rev. A {\bf 49}, 3022, (1994).
\bibitem{yue} H. P. Yuen in {\em Squeezed States and Uncertainty 
Relations}, D. Han et al Eds. NASA CP 3135, p 13 (1992). 
\end{thebibliography}
\end{document}